\documentclass[aps,preprint,showpacs]{revtex4}
\usepackage{graphicx}

\begin{document}

\title{Biased random satisfiability problems: From easy to hard instances}

\author{A. Ramezanpour}
\email{ramezanpour@iasbs.ac.ir}

 \affiliation{Institute for Advanced Studies in Basic Sciences,
Zanjan 45195-1159, Iran}

\author{S. Moghimi-Araghi}
\email{samanimi@sharif.edu}

\affiliation{Department of Physics, Sharif University of
Technology, P.O.Box 11365-9161, Tehran, Iran}

\date{\today}

\begin{abstract}
In this paper we study biased random K-SAT problems in which each
logical variable is negated with probability $p$. This
generalization provides us a crossover from easy to hard problems
and would help us in a better understanding of the typical
complexity of random K-SAT problems. The exact solution of 1-SAT
case is given. The critical point of K-SAT problems and results of
replica method are derived in the replica symmetry framework. It
is found that in this approximation $\alpha_c \propto p^{-(K-1)}$
for $p\rightarrow 0$. Solving numerically the survey propagation
equations for $K=3$ we find that for $p<p^* \sim 0.17$ there is no
replica symmetry breaking and still the SAT-UNSAT transition is
discontinuous.
\end{abstract}

 \maketitle

\section{Introduction}\label{1}
Optimization problems are subject of recent studies in the context
of complex systems \cite{ya,ks,mzkst,m1}. Random K-SAT problems
are well known examples of these problems which have their origin
in computer science and complexity theory
\cite{p,m2,p1,mmz,mz0,mz,mez}. Finding the configuration of $N$
logical variables which satisfies a formula of $M$ clauses is a
hard problem and indeed lies in
the class of NP-complete problems for $K\ge 3$\cite{c}.\\
From a physical point of view the interesting feature of random
K-SAT problems is the presence of phase transitions in the
thermodynamic limit where $N$ and $M$ approach infinity and
$\alpha:=M/N$ remains finite \cite{ks}. Here the transition is
between SAT and UNSAT phases where a typical instance of the
formula is satisfied or unsatisfied respectively with probability
$1$. It is around the phase transition that the time needed to
find the solution of a typical instance grows exponentially with
the size of the problem $N$.\\
In this paper we study a generalized version of the random K-SAT
problems where each logical variable is negated with probability
$p$ rather than $1/2$ as in the original random K-SAT. The aim of
this generalization is to go continuously from easy instances of
the problem to hard ones. Clearly for $p=0$ we have an easy random
K-SAT for all values of $K$. On the other hand as the studies
indicate, the problem is hard for $p=1/2$ and $K\ge 3$. Thus one
expects a crossover from easy to hard region by increasing $p$
from zero. Is it possible to define a point beyond which one can
say that the problem becomes hard? How the problem approaches the
hard regime? What are the universal features of this crossover?
These are some questions which can provide us a deeper
understanding of the typical complexity of random K-SAT
problems.\\
A similar problem to one that we are going to study is the
Horn-SAT problem \cite{w} where all the clauses have at least one
negated variable. It is an easy problem and solved in a polynomial
time. However, notice that as long as $K$ is small fluctuations
paly an important role in our problem and this could give rise to
significantly different behaviors for the problem. There is also
another problem called $2+p-SAT$ \cite{mzkst} which by tuning $p$
goes from a $2-SAT$ to a $3-SAT$ problem. It is close to what we
like to do in this paper but here we are able to study the
easy-hard crossover for general $K$ and it is a more general
problem to this end.\\
In the following we first give the exact solution of 1-SAT problem
by a statistical mechanics approach. We find the average number of
unsatisfied clauses and the average number of solutions in the
ground state of the system and explain the origin of their
behaviors. Utilizing the cavity method and assuming the replica
symmetry, we derive a relation for the critical point of K-SAT
problems. It is found that in general $\alpha_c \propto
p^{-(K-1)}$ as $p \rightarrow 0$.  Next we obtain the free energy
and the distribution of effective fields with the aid of replica
method and in the replica symmetry approximation. Finally we
resort to the numerical solution of survey propagation
equations\cite{bmz} for the case of $K=3$ and compare the
extracted critical points with the predictions of replica symmetry
assumption. It is found that for $p>p^*\sim 0.17$ the replica
symmetry breaks at some point $\alpha_d< \alpha_c$ whereas for
$p<p^*$ we are always in the easy -SAT phase if
$\alpha < \alpha_c$.\\

The paper is organized as follows. In the next section we define
the problem. Section \ref{3} is devoted to the study of 1-SAT
problem. Assuming the presence of replica symmetry we give the
results of cavity and replica methods in sections \ref{4} and
\ref{5}. Survey propagation equations for the case $K=3$ are
numerically studied in section \ref{6}. Section \ref{7} includes
the conclusion remarks of the paper.

\section{The problem definition}\label{2}
\begin{figure}
\includegraphics[width=8cm]{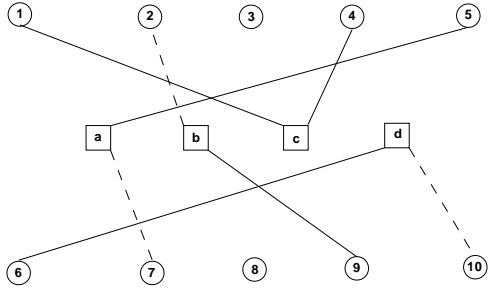}
\caption{Factor graph of the formula given in Eq.\ref{for}.
}\label{f1}
\end{figure}

 We take $N$ logical variables $\{x_i|i=1,\ldots,N\}$ where $x_i=1$
 if the corresponding variable is true and otherwise $x_i=0$.
 Alternatively we can speak of $N$ Ising variables $S_i:=2x_i-1$.
 On the other side we have a formula which consist of $M$ clauses which have been joined
 to each other by logical AND. Each clause in turn contains $K$
 logical variable selected randomly from the list of our $N$
 variables. These variables, which join to each other by logical
 OR, are negated with probability $p$. One obtains
 the original random K-SAT problem by choosing $p=1/2$. Here there is an example of
 a 2-SAT formula with $4$ clauses and $10$ logical variables
 \begin{equation}\label{for}
 F:=(x_5\vee \overline{x}_7)\wedge(\overline{x}_2\vee x_9)\wedge(x_1\vee x_4)
 \wedge(x_6\vee \overline{x}_{10}).
 \end{equation}
A very useful concept in these problems is the factor graph, a
bipartite graph of variable nodes and function nodes. In
Fig.\ref{f1} we have shown the factor graph of formula given
above. In this figure the logical variables and the clauses have
been represented by the circles (variable nodes) and the squares
(function nodes) respectively. An edge in this graph only connects
a variable node to a function node and its style gives the nature
of the logical variable in the associated clause. Here dashed
edges are used to indicate that the negated variable enters the
clause. We can summarize the factor graph in matrix $C_{M\times
N}$ with elements $C_{a,i} \in \{0, +1, -1\}$. In fact $C_{a,i}$
is $+1$ or $-1$ if clause $a$ contains variable $i$ or its negated
respectively. Otherwise $C_{a,i}=0$.
\section{The simple case of $K=1$}\label{3}
\begin{figure}
\includegraphics[width=8cm,angle=270]{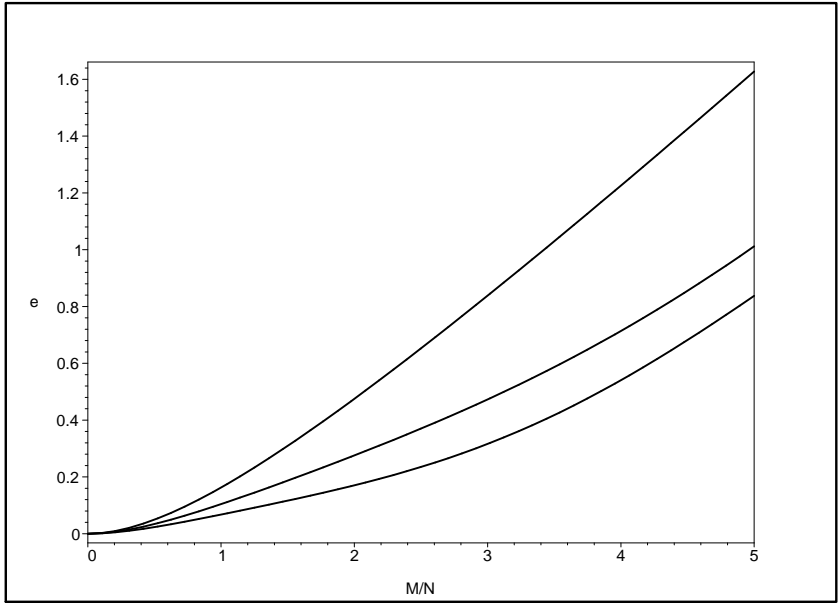}
\caption{Ground state energy of 1-SAT problem from top to bottom
for $p=1/2$, $p=1/4$ and $p=1/6$. }\label{f2}
\end{figure}

\begin{figure}
\includegraphics[width=8cm,angle=270]{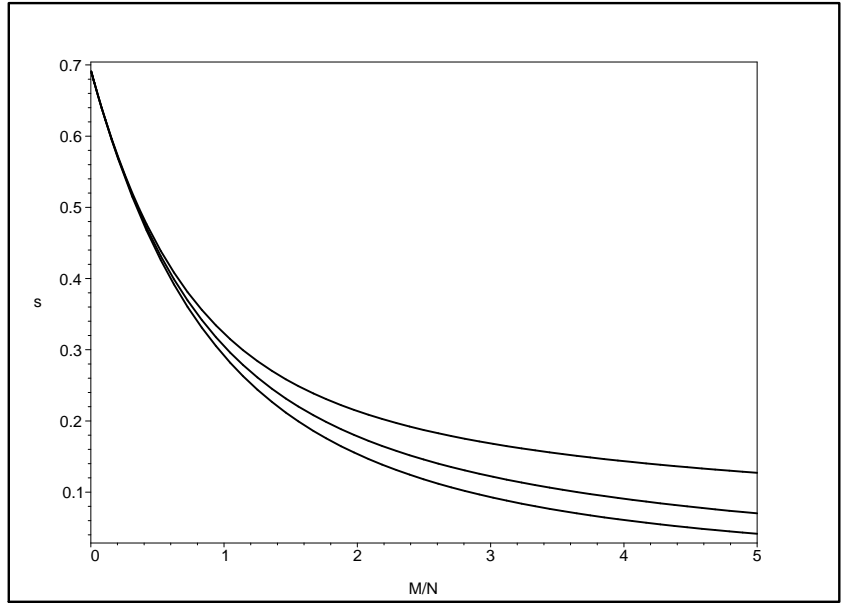}
\caption{Ground state entropy of 1-SAT problem from top to bottom
for $p=1/2$, $p=1/4$ and $p=1/6$. }\label{f3}
\end{figure}

We start by giving the exact behavior of 1-SAT problem by keeping
a statistical mechanics approach \cite{mz0,mz}. We define the
energy of a formula as the number of violated clauses, that is
\begin{equation}
E[S,C]:=\sum_{a=1}^M [1-\sum_{i=1}^N C_{a,i}S_i]/2,
 \end{equation}
 where $S$ denotes the configuration
 of Ising variables. Note that by definition
\begin{equation}
\sum_{a=1}^MC_{a,i}=t_i-f_i,
 \end{equation}
where $t_i$ and $f_i$ give the number of full lines and dashed
lines respectively emanating from variable node $i$. The set of
$\{t_i,f_i\}$ only depends on the structure of the factor graph.
Utilizing the above facts and summing over spin configurations the
partition function reads
\begin{equation}
Z[C]:=\sum_{S} e^{-\beta E[S,C]}=\prod_i\left( e^{-\beta
t_i}+e^{-\beta f_i}\right),
\end{equation}
where $\beta=1/k_BT$. The free energy per variable, $-(1/\beta
N)\ln(Z[C])$, still depends on the structure of the
 factor graph and we should take an average over this kind of
 disorder.\\
 The probability to have the set $\{t_i,f_i\}$ is given by
 \begin{equation}
P[\{t,f\}]=\left(M!/N^M\right) \prod_i\left(
\frac{p^{f_i}(1-p)^{t_i}}{(t_i!f_i!)}\right).
\end{equation}
Then the averaged free energy in the thermodynamic limit reads
\begin{equation}\label{f}
f:=\overline{f[C]}=\alpha/2-1/\beta\{\ln(2)+e^{-\alpha}
\sum_{n,m=-\infty}^{\infty}
J_m(q\alpha)I_n(\alpha)\ln(\cosh(\beta(m+n)/2))\},
\end{equation}
where we have defined $q:=1-2p$. Moreover $J_m(\alpha)$ and
$I_n(\alpha)$ are the Bessel functions of first kind. Now from
Eq.(\ref{f}) one can easily find the average energy per variable
\begin{equation}
e:=\overline{<E>}/N=\alpha/2-e^{-\alpha}\sum_{m,n=-\infty}^{\infty}
J_m(q\alpha)I_n(\alpha)(m+n)/2 \tanh(\beta(m+n)/2).
\end{equation}
In the same way the entropy per variable is given by
\begin{eqnarray}\label{s}
s:=\overline{<S>}/N=\ln(2)+ e^{-\alpha}\sum_{m,n=-\infty}^{\infty}
J_m(q\alpha)I_n(\alpha)\{\ln(\cosh(\beta(m+n)/2))\\ \nonumber
-\beta(m+n)/2 \tanh(\beta(m+n)/2)\}.
\end{eqnarray}
We are interested in the ground state properties of the problem
and to this end we need to take the limit $\beta \rightarrow
\infty$ in the above relations. After some simplifications we find
for the ground state energy
\begin{equation}
e_G=(1-p)\alpha-e^{-\alpha}\sum_{m+n>0}
(m+n)J_m(q\alpha)I_n(\alpha).
\end{equation}
In Fig.\ref{f2} we have shown the behavior of this quantity versus
$\alpha$ and for some values of $p$. As expected the problem is
always in the UNSAT phase where an infinite number of clauses are
not satisfied in the thermodynamic limit. Indeed the probability
that a new clause is not satisfied is given by $2p(1-p)\alpha$ and
this quantity has always a nonzero value as long as $p$ and
$\alpha$ are nonzero.\\
The entropy of system at zero temperature is found from
Eq.(\ref{s})
\begin{equation}
s_G=e^{-\alpha}I_0(\alpha\sqrt{1-q^2})\ln(2).
\end{equation}
We have numerically computed this quantity and have shown its
behavior with $\alpha$ in Fig.\ref{f3} . Clearly the number of
solutions is the same for different values of $p$ and small
$\alpha$. It is due to the fact that for small $\alpha$ each
clause constrain a variable regardless of the nature of variable
in that clause. However for large $\alpha$ the situation is
different. In fact when a variable contributes in different forms
in a number of clauses, with a larger probability it can take the
values $0,1$ without changing the number of violated clauses. This
is the reason for the smaller entropy of smaller values of $p$ in
Fig.\ref{f3}.

\section{Cavity method }\label{4}
\begin{figure}
\includegraphics[width=8cm]{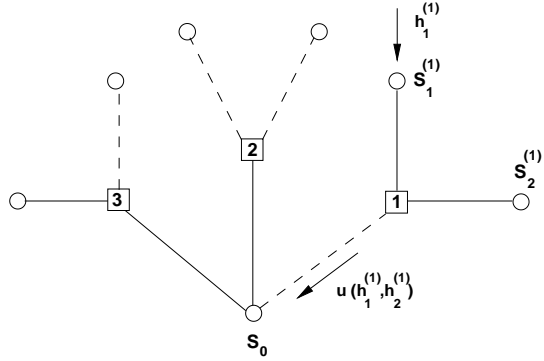}
\caption{$S_0$ experiences a cavity filed determined from the
fields experienced by the other neighbors of function nodes
$1,2,3$.}\label{f4}
\end{figure}

\begin{figure}
\includegraphics[width=8cm]{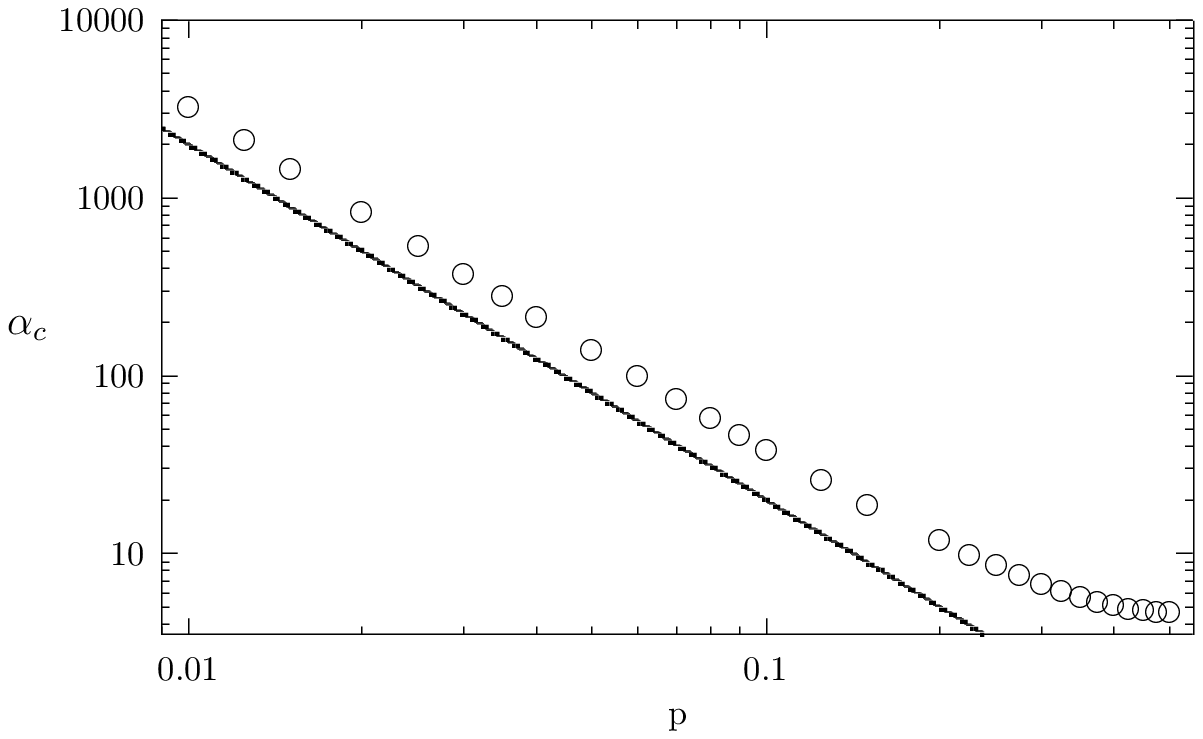}
\caption{$\alpha_c$ versus $p$ for $K=3$ and in the replica
symmetric approximation. The line shows a power law of exponent
$2$ }\label{f5}
\end{figure}

In this section we apply the cavity method \cite{mez} to our
problem. For simplicity we work only in the
replica symmetric scheme and zero temperature.\\
Consider the variable node $S_0$ with $z=3$ neighbors in
Fig.\ref{f4}.
It is known that the cavity field experienced by variable $S_0$ is
\begin{equation}
h_0=\sum_{a=1}^z u(h_1^{(a)},\ldots,h_{K-1}^{(a)}),
\end{equation}
where
\begin{equation}\label{u}
u(h_1,\ldots,h_{K-1})=C_0\prod_{r=1}^{K-1}\theta(-C_rh_r).
\end{equation}
Here $\theta(x)$ is the known step function which is $1$ for $x>0$
and $0$ otherwise. Note that $u$ and $h$ are stochastic
quantities. In the replica symmetric framework the probability
distribution of these quantities are given by \cite{mez}
\begin{eqnarray}
Q(u)=c_0\delta(u)+c_-\delta(u+1)+c_+\delta(u-1),\\ \nonumber
P(h)=\sum_{l=-\infty}^{-1}P_-(h)\delta(l-h)+P_0\delta(h)+
\sum_{l=1}^{\infty}P_+(h)\delta(l-h).
\end{eqnarray}
Then it is an easy exercise to show that for $h\ge 0$ the
probability distribution of $h$ is
\begin{equation}
P_+(h)=\sum_{z=h}^{\infty}f_{K\alpha}(z)\sum_{r=0}^{[(z-h)/2]}
\left(\begin{array}{c}
  z \\
  2r+h \\
\end{array}\right)\left(\begin{array}{c}
  2r+h \\
  r \\
\end{array}\right)c_0^{z-2r-h}c_-^rc_+^{r+h}.
\end{equation}
Here $[\ldots]$ denotes the integer part of enclosed quantity and
$f_{K\alpha}(z)$ is the degree distribution of variable nodes
\begin{equation}\label{fz}
f_{K\alpha}(z)=\left(z^{K\alpha}/z!\right)e^{-K\alpha}.
\end{equation}
Similarly for $h\le 0$, $P_-(-h)$ is obtained by interchanging the
role of $c_-$ and $c_+$ in the above equation. Simplifying the
above relations one finds that
\begin{equation}\label{p0}
P_0=e^{-K\alpha(1-c_0)}I_0(2K\alpha\sqrt{c_-c_+}),
\end{equation}
and after some straightforward algebra for
$P_+=\sum_{h=1}^{\infty}P_+(h)$ one obtains
\begin{equation}\label{p00}
P_++P_0=e^{-K\alpha c_+}\int_{K\alpha
c_-}^{\infty}e^{-t}I_0(2\sqrt{K\alpha c_+t}).
\end{equation}
Eqs.(\ref{p0}) and (\ref{p00}) are two independent relations which
along with the normalization condition $P_-+P_0+P_+=1$ determine
$P(h)$ and $Q(u)$. However we still need to derive the relations
between $\{c_0,c_-,c_+\}$ and $\{P_0,P_-,P_+\}$. To this end note
that from Eq.(\ref{u}) one has
\begin{equation}\label{cpr}
c_0=1-[pP_++(1-p)P_-]^{K-1}:=1-c_p,\hskip 1cm c_-=pc_p, \hskip 1cm
c_+=(1-p)c_p.
\end{equation}
Summing the above relations one can use Eqs.(\ref{p0}) and
(\ref{p00}) to find the following equation for $c_p$
\begin{equation}\label{cp}
c_p=[g(c_p)]^{K-1},
\end{equation}
where
\begin{equation}\label{gcp}
g(c_p):=1-p-(1-p)P_0-(1-2p)P_+.
\end{equation}
For $p=1/2$ we recover the known relation for the effective filed
distribution
\begin{equation}\label{php12}
P(h)=e^{-K\alpha c_p}I_h(K\alpha c_p),
\end{equation}
where now $c_p$ satisfies
\begin{equation}\label{cp12}
c_p=[\frac{1-e^{-K\alpha c_P}I_0(K\alpha c_p)}{2}]^{K-1}.
\end{equation}
Returning to our general problem we find from Eq.(\ref{gcp}) that
$g(0)=0$ and
\begin{equation}
\frac{dg(c_p)}{dc_p}|_{c_p=0}=2K\alpha p(1-p).
\end{equation}
Thus as expected Eq.(\ref{cp}) suggests a continues transition for
$K=2$ and discontinuous transitions for $K\ge 3$. Indeed for $K=2$
the critical value of $\alpha$ is given by
\begin{equation}
\alpha_c=\frac{1}{4p(1-p)}.
\end{equation}
Due to the absence of replica symmetry breaking the above results
are exact for $K=2$. As expected for $p=1/2$ we get the known
value of $\alpha_c=1$ and $\alpha_c
\rightarrow \infty$ for $p\rightarrow 0,1$.\\
What can be said about $\alpha_c$ for general $K$? Let us focus on
the behavior of $\alpha_c$ as $p\rightarrow 0$. First note that
just above the critical point $P_+$ takes a finite value and from
the definition of $c_p$, Eq.(\ref{cpr}), one obtains $c_p \propto
p^{K-1}$. On the other hand in $g(c_p)$, $c_p$ always appears
along with $\alpha$ as $\alpha c_p$. Expanding $g(c_p)$ one finds
that only for a finite $\alpha c_p$ the two sides of Eq.(\ref{cp})
would have the same scaling with $p$. This in turn suggests that
the critical value of $\alpha$ should scale as $p^{-(K-1)}$.\\In
Fig.\ref{f5} we have solved Eq.(\ref{cp}) numerically for $K=3$.
Indeed for $K\ge3$ the replica symmetric predictions provide an
upper bound for $\alpha_c$ \cite{fl}. For $K=3$ we found that as
expected $\alpha_c$ approaches to infinity like $p^{-2}$ as
$p\rightarrow 0$.

\section{Replica approach}\label{5}
In the following we will keep the same lines as Ref.\cite{mz} to
calculate the free energy of biased random K-SAT problems in the
replica formalism. As before we can write energy or the number of
violated clauses as
\begin{equation}
E[S,C]=\sum_{l=1}^M\delta\left(\sum_{i=1}^N C_{l,i}S_i+K\right).
\end{equation}
Our goal is to find $\overline{\ln(Z[C])}$ and to this end we need
to obtain
\begin{equation}
Z_n:=\overline{Z[C]^n}=\sum_{S^1,\ldots,S^n}\overline{e^{-\beta
\sum_{a=1}^n E[S^a,C]}}.
\end{equation}
The overline denotes averaging with respect to the random
structure of the factor graph. Due to the independent nature of
clauses on can write
\begin{equation}\label{zn}
Z_n=\sum_{S^1,\ldots,S^n}\zeta^M,
\end{equation}
where
\begin{equation}
\zeta=\overline{e^{-\beta \sum_a \delta\left(\sum_i
C_{l,i}S_i^a+K\right)}}.
\end{equation}
Then for a single clause we can use the fact that
\begin{equation}
 \delta\left(\sum_i C_{l,i}S_i+K\right)=\prod_{j=1}^K\delta\left(S_{i_j}^a+C_j\right),
\end{equation}
to write the following expression for $\zeta$
\begin{equation}
\zeta=\frac{1}{\left(\begin{array}{c}N \\ K \\ \end{array}\right)}
\sum_{C_1,\ldots,C_K=\pm
1}p^{\nu[C]}(1-p)^{K-\nu[C]}\sum_{i_1,\ldots,i_K=1}^N e^{-\beta
\sum_a \prod_{j=1}^K \delta\left(S_{i_j}^a+C_j\right)}(1+O(1/N)+
\ldots ),
\end{equation}
where $\nu[C]$ is the number of minus elements in the set
$\{C_j|j=1,\ldots,K\}$. Let us also define
\begin{equation}
x(\vec{\sigma}):=1/N\sum_{i=1}^N\delta\left(\vec{S}_i-\vec{\sigma}\right),
\end{equation}
where $\vec{\sigma}$ and $\vec{S}_i$ are vectors of $n$ Ising
elements in the replica space. Then apart from some irrelevant
terms and constants we obtain
\begin{equation}
\zeta=\sum_{C_1,\ldots,C_K=\pm 1}p^{\nu[C]}(1-p)^{K-\nu[C]}
\sum_{\vec{\sigma}_1,\ldots,\vec{\sigma}_K}x(-C_1\vec{\sigma}_1)\ldots
x(\vec{-C_K\sigma}_K)e^{-\beta\sum_a \prod_{j=1}^K
\delta\left(\sigma_j^a-1\right)}.
\end{equation}
and thus
\begin{equation}\label{tf}
Z_n \sim \sum_{x(\vec{\sigma})}e^{-N\tilde{f}(x)}, \hskip 1cm
\tilde{f}(x):=-\alpha
\ln(W(x))+\sum_{\vec{\sigma}}x(\vec{\sigma})\ln(x(\vec{\sigma})),
\end{equation}
where $\alpha:=M/N$ and
\begin{equation}
W(x):=\sum_{\nu|K}p^{\nu}(1-p)^{K-\nu}\sum_{\vec{\sigma}_1,\ldots,\vec{\sigma}_K}
\prod_{j=1}^{\nu}x(\vec{\sigma}_j)\prod_{j=\nu+1}^Kx(-\vec{\sigma}_j)
e^{-\beta \sum_a \prod_{j=1}^K\delta\left(\sigma_j^a-1\right)}.
\end{equation}
In this equation $\sum_{\nu|K}$ means a sum over all the
selections of $\nu$ variables from $K$ ones and in the same time
it orders these selected variables in the beginning of a
$K$-member list.\\
Now we should find a form for $x(\vec{\sigma})$ which minimizes
$\tilde{f}(x)$. As in previous studies \cite{mz} we use the
following ansatz in the replica symmetric scheme
\begin{equation}
x(\vec{\sigma})=\int_{-1}^{1}dmP(m)\prod_{a=1}^n\left(\frac{1+m\sigma^a}{2}\right).
\end{equation}
Note that for $p\ne 1/2$ we do not have the symmetry relation
$x(-\vec{\sigma})=x(\vec{\sigma})$ and so $P(m)$ is not an
even function.\\
Applying the above ansatz we find
\begin{equation}\label{dx}
W(x)=\sum_{\nu=0}^K B(\nu,K;p)\int\prod_{j=1}^K dm_jP(m_j)\prod_a
A_{\nu,K-\nu}(m),
\end{equation}
where $B(\nu,K;p)$ is the binomial distribution
\begin{equation}
B(\nu,K;p)=\left(\begin{array}{c}K \\ \nu \\
\end{array}\right) p^{\nu}(1-p)^{K-\nu},
\end{equation}
and
\begin{equation}
A_{\nu,K-\nu}(m):=\sum_{\sigma_1^a,\ldots,\sigma_K^a}\prod_{j=1}^{\nu}
\left(\frac{1+m_j\sigma_j^a}{2}\right)\prod_{j=\nu+1}^K\left(\frac{1-m_j\sigma_j^a}{2}\right)
e^{-\beta\prod_{j=1}^K\delta\left(\sigma_j^a-1\right)},
\end{equation}
Doing the sum over $\sigma$ one obtains
\begin{equation}
A_{\nu,K-\nu}(m)=1+\left(e^{-\beta}-1\right)\prod_{j=1}^{\nu}
\left(\frac{1+m_j}{2}\right)\prod_{j=\nu+1}^K\left(\frac{1-m_j}{2}\right).
\end{equation}
If we Optimize $\tilde{f}(x)$ with respect to $x(\vec{\sigma})$ we
find that
\begin{equation}
x(\vec{\sigma})=\Lambda e^{\alpha W'(x)/W(x)},
\end{equation}
where
\begin{equation}
W'(x):=\frac{\delta}{\delta x(\vec{\sigma})}W(x),
\end{equation}
and $\Lambda$ is determined from the normalization condition.
After some calculations  one finds the following relation for
$W'(x)$
\begin{equation}\label{dx'}
W'(x)=\sum_{\nu=0}^K B(\nu,K;p)\int \prod_{j=1}^K dm_jP(m_j)[\nu
\prod_{a_+}A_{\nu-1,K-\nu}(m)+(K-\nu)\prod_{a_-}A_{\nu,K-\nu-1}(m)],
\end{equation}
where $\prod_{a_{\pm}}$ denotes a product over the indices $a$ for
them $\sigma^a=\pm 1$. Moreover in this equation
\begin{eqnarray}\label{bb}
A_{\nu-1,K-\nu}(m)=1+\left(e^{-\beta}-1\right)\prod_{j=1}^{\nu-1}
\left(\frac{1+m_j}{2}\right)\prod_{j=\nu}^{K-1}\left(\frac{1-m_j}{2}\right), \\
\nonumber
A_{\nu,K-\nu-1}(m)=1+\left(e^{-\beta}-1\right)\prod_{j=1}^{\nu}
\left(\frac{1+m_j}{2}\right)\prod_{j=\nu+1}^{K-1}\left(\frac{1-m_j}{2}\right).
\end{eqnarray}
 We are interested to limit $n\rightarrow 0$ where $x(\vec{\sigma})$
can be written as
\begin{equation}
x(\vec{\sigma})=\int dmP(m)e^{u \ln\left(\frac{1+m}{1-m}\right)}.
\end{equation}
 Now doing the standard algebra \cite{mz} we
 find the following self consistency relation for $P(m)$
\begin{equation}\label{pmf}
P(m)=\frac{2}{1-m^2}\int_{-\infty}^{\infty} du
e^{-iu\ln\left(\frac{1+m}{1-m}\right)} e^{-\alpha K+\alpha
W'(iu)},
\end{equation}
where
\begin{eqnarray}\label{dxx}
W'(u)=\sum_{\nu=0}^{K-1} B(\nu,K;p)\int \prod_{j=1}^{K-1}
dm_jP(m_j)[\nu e^{u\ln\left(A_{\nu-1,K-\nu}(m)\right)}\\
\nonumber +(K-\nu)e^{-u\ln\left(A_{\nu,K-\nu-1}(m)\right)}].
\end{eqnarray}
Similarly for the free energy we find
\begin{eqnarray}\label{ff}
\beta f=-\ln(2)-\alpha(1-K) \sum_{\nu=0}^K B(\nu,K;p)\int
\prod_{j=1}^K dm_jP(m_j) \ln\left(A_{\nu,K-\nu}\right) \\
\nonumber -\alpha/2 \sum_{\nu=0}^K B(\nu,K;p)\int
\prod_{j=1}^{K-1} dm_jP(m_j) [\nu \ln\left(A_{\nu-1,K-\nu}\right) \\
\nonumber +(K-\nu) \ln\left(A_{\nu,K-\nu-1}\right)]+ 1/2\int dm
P(m)\ln\left(1-m^2\right).
\end{eqnarray}
Eqs.(\ref{ff}) and (\ref{pmf}) return the known relations for
$p=1/2$ \cite{mz} when $P(m)$ is an even function. Finally let us
consider the limit $\beta \rightarrow \infty$ of Eqs.(\ref{pmf})
and (\ref{ff}). To this end we should work with effective fields
$z$ given by $m=\tanh(\beta z/2)$ \cite{mz0}. Then for
$\beta=\infty$ we get
\begin{equation}
R(z)=e^{-\alpha K} \int_{-\infty}^{\infty}du e^{-iuz}e^{\alpha
W'(iu)}.
\end{equation}
Now the relation that gives $W'(u)$ reads
\begin{eqnarray}
W'(u)=K-K[p\int_0^{\infty}dzR(z)+(1-p)\int_{-\infty}^0dzR(z)]^{K-1}
\\ \nonumber+\sum_{\nu=0}^{K-1} B(\nu,K;p)
[\nu \int D_{\nu-1,K-\nu}e^{-u \hskip0.1cm\textrm{min}(1,z_1,\ldots,z_{\nu-1},-z_{\nu},\ldots,-z_{K-1})}\\
\nonumber +(K-\nu)\int D_{\nu,K-\nu-1}e^{u\hskip0.1cm
\textrm{min}(1,z_1,\ldots,z_{\nu},-z_{\nu+1},\ldots,-z_{K-1})}].
\end{eqnarray}
where
\begin{eqnarray}
D_{\nu-1,K-\nu}:=\int_{0}^{\infty}\prod_{j=1}^{\nu-1}dz_jR(z_j)\int_{-\infty}^{0}\prod_{j=\nu}^{K-1}dz_jR(z_j),
\\ \nonumber D_{\nu,K-\nu-1}:=\int_{0}^{\infty}\prod_{j=1}^{\nu}dz_jR(z_j)\int_{-\infty}^{0}\prod_{j=\nu+1}^{K-1}dz_jR(z_j).
\end{eqnarray}
For the free energy in this limit we have
\begin{eqnarray}
f=\alpha(1-K) \sum_{\nu=0}^K B(\nu,K;p) \int
D_{\nu,K-\nu}\hskip0.1cm
\textrm{min}(1,z_1,\ldots,z_{\nu},-z_{\nu+1},\ldots,-z_{K}) \\
\nonumber +\alpha/2 \sum_{\nu=0}^K B(\nu,K;p) [\nu \int
D_{\nu-1,K-\nu}\hskip0.1cm
\textrm{min}(1,z_1,\ldots,z_{\nu-1},-z_{\nu},\ldots,-z_{K-1})
\\ \nonumber +(K-\nu)\int D_{\nu,K-\nu-1}\hskip0.1cm
\textrm{min}(1,z_1,\ldots,z_{\nu},-z_{\nu+1},\ldots,-z_{K-1})]\\
\nonumber +1/2[\int_{-\infty}^{0}
dzR(z)z-\int_{0}^{\infty}dzR(z)z].
\end{eqnarray}
Considering the simple case of $K=1$ the effective field
distribution reads
\begin{equation}
R(z)=e^{-\alpha}\sum_{m=-\infty}^{\infty}\left(I_0(q\alpha)I_m(\alpha)+\sum_{n=1}^{\infty}
(-1)^nI_{2n}(q\alpha)[I_{2n-m}(\alpha)+I_{2n+m}(\alpha)]\right)\delta(z-m),
\end{equation}
which for $p=1/2$ returns
\begin{equation}
R(z)=e^{-\alpha}\sum_{n=-\infty}^{\infty}I_n(\alpha)\delta_{z-n}.
\end{equation}
Compare the above relation with Eq. \ref{php12} which gives the
effective filed distribution in the cavity method and in the
replica symmetric approximation. In fact the two distributions are
the same as they should be as long as we use an ansatz in which
the effective fields take integer values.

\section{Survey propagation equations}\label{6}
\begin{figure}
\includegraphics[width=8cm]{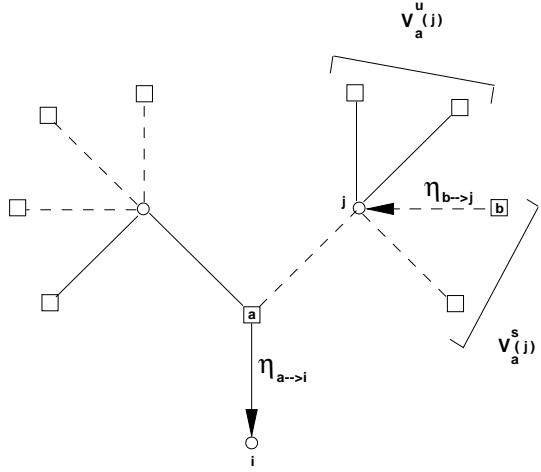}
\caption{The survey warning $\eta_{a\rightarrow i}$ is determined
by the set of surveys $\eta_{b\rightarrow j}$. }\label{f6}
\end{figure}

In this section we study the behavior of 3-SAT problem by means of
numerical solution of survey propagation equations
\cite{bmz,bmwz}. Let us first write the general form of these
equations. We define $\eta_{a\rightarrow i}$ as the probability
that in a state selected randomly from the existing states of the
problem, the clause $a$ sends a warning to variable $i$ to take
the value that satisfies it. This warning is sent if the other
members of $a$ do not satisfy this clause. We denote by $V(a)$ the
set of neighbors of $a$. Then assuming a tree structure for the
factor graph we have
\begin{equation}\label{eta0}
\eta_{a\rightarrow i}=\prod_{j\in V(a)|i}P_a^u(j),
\end{equation}
where the product is over all the neighbors of $a$ excluding $i$
and $P_a^u(j)$ is the probability that variable $j$ does not
satisfy clause $a$. Let us denote by $V_a^s(j)$ the set of clauses
that variable $j$ appears in them as it appears in clause $a$,
Fig.\ref{f6}.
The remaining set of clauses are denoted by $V_a^u(j)$. With these
definitions $P_a^u(j)$ is given by \cite{bmz}
\begin{equation}\label{Pauj}
P_a^u(j)=\frac{\Pi_{j\rightarrow a}^u}{\Pi_{j\rightarrow
a}^s+\Pi_{j\rightarrow a}^0+\Pi_{j\rightarrow a}^u},
\end{equation}
where
\begin{eqnarray}\label{Pii}
\Pi_{j\rightarrow a}^u=[1-\prod_{b\in
V_a^u(j)}\left(1-\eta_{b\rightarrow j}\right)]\prod_{b\in V_a^s(j)}\left(1-\eta_{b\rightarrow j}\right),\\
\nonumber \Pi_{j\rightarrow a}^s=[1-\prod_{b\in
V_a^s(j)}\left(1-\eta_{b\rightarrow j}\right)]\prod_{b\in V_a^u(j)}\left(1-\eta_{b\rightarrow j}\right),\\
\nonumber \Pi_{j\rightarrow a}^0=\prod_{b\in
V(j)|a}\left(1-\eta_{b\rightarrow j}\right).
\end{eqnarray}
Now starting from an arbitrary configuration for the warnings sent
along the edges of the factor graph one obtains the new values of
$\eta$'s from Eqs.(\ref{eta0},\ref{Pauj},\ref{Pii}) and repeat
this procedure until reaches to a stationary state. It is believed
that in the whole region of SAT phase the above equations result
in the correct solutions of random K-SAT problems \cite{bmwz}.
Here we apply the same procedure to 3-SAT problem to compute
$\Sigma$, the complexity of our problems. The complexity of a
formula is the logarithm of the number of states and reads
\cite{bmz}
\begin{equation}\label{comp}
\Sigma=\sum_{a=1}^M\Sigma_a-\sum_{i=1}^N(z_i-1)\Sigma_i,
\end{equation}
where
\begin{eqnarray}\label{compai}
\Sigma_a=\log[\prod_{j\in V(a)}\left(\Pi_{j\rightarrow
a}^s+\Pi_{j\rightarrow a}^0+
\Pi_{j\rightarrow a}^u\right)-\prod_{j\in V(a)}\Pi_{j\rightarrow a}^u],\\
\nonumber \Sigma_i=\log[\Pi_i^-+\Pi_i^0+\Pi_i^+],
\end{eqnarray}
and
\begin{eqnarray}\label{Pi}
\Pi_{i}^-=[1-\prod_{a\in V_-(i)}(1-\eta_{a\rightarrow
i})]\prod_{a\in V_+(i)}(1-\eta_{a\rightarrow i}),\\ \nonumber
\Pi_{i}^+=[1-\prod_{a\in
V_+(i)}(1-\eta_{a\rightarrow i})]\prod_{a\in V_-(i)}(1-\eta_{a\rightarrow i}),\\
\nonumber \Pi_{i}^0=\prod_{a\in V(i)}(1-\eta_{a\rightarrow i}).
\end{eqnarray}
In these equations $V(i)$ denotes the set of $z_i$ neighbors of
variable node $i$, $V_+(i)$ is the set of function nodes in $V(i)$
that have been connected to $i$ by a full line and $V_-(i)$ gives
the
complementary subset.\\
It is known that $\Sigma$ is zero in the replica symmetric and
UNSAT phases and nonzero in the hard-SAT phase \cite{bmz}.
Increasing $\alpha$ one first encounters the replica symmetry
breaking point at $\alpha_d$ where $\Sigma$ takes discontinuously
its maximum value $\Sigma_m$. After this stage $\Sigma$ decreases
and finally vanishes at the critical point $\alpha_c$. One can use
these properties of $\Sigma$ to compute  $\alpha_d$ and
$\alpha_c$.\\ To solve the survey propagation equations we used
the software given in \cite{ictp}.
\begin{figure}
\includegraphics[width=8cm]{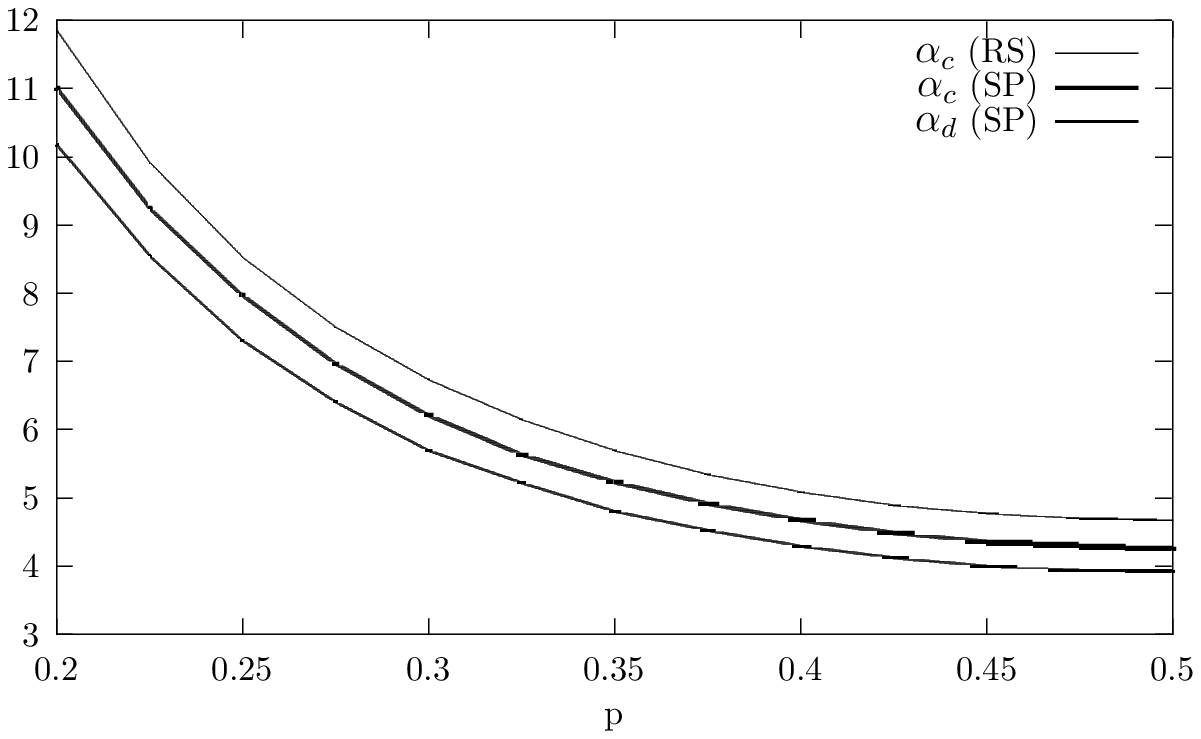}
\caption{From top to bottom: the replica symmetry predictions for
$\alpha_c$ (RS), survey propagation predictions of $\alpha_c$ (SP)
and $\alpha_d$ (SP) for $K=3$ and $N=10000$. The numerical results
have been obtained for one realization with the convergence limit
equal to $0.001$. }\label{f7}
\end{figure}
In Fig.\ref{f7} we have shown the results of this computation for
$\alpha_c$ and $\alpha_d$ and compared $\alpha_c$ with the
predictions of replica symmetric case. As the figure shows the
behavior of $\alpha_c$ with $p$ is qualitatively similar to the
one obtained with the replica symmetry assumption. The represented
data have been restricted to relatively large values of $p$. It is
due to the fact that for smaller values of $p$ the complexity
vanishes and we are not able to identify $\alpha_c$ by looking at
$\Sigma$.\\ In Fig.\ref{f8} we also showed the behavior of
$\Sigma_m$ versus $p$. It is seen that around $p^*=0.17$ the
maximum complexity vanishes discontinuously. Then it can be
concluded that for $p<p^*$ we have a simple problem as in the
regime of easy-SAT phase.
\begin{figure}
\includegraphics[width=8cm]{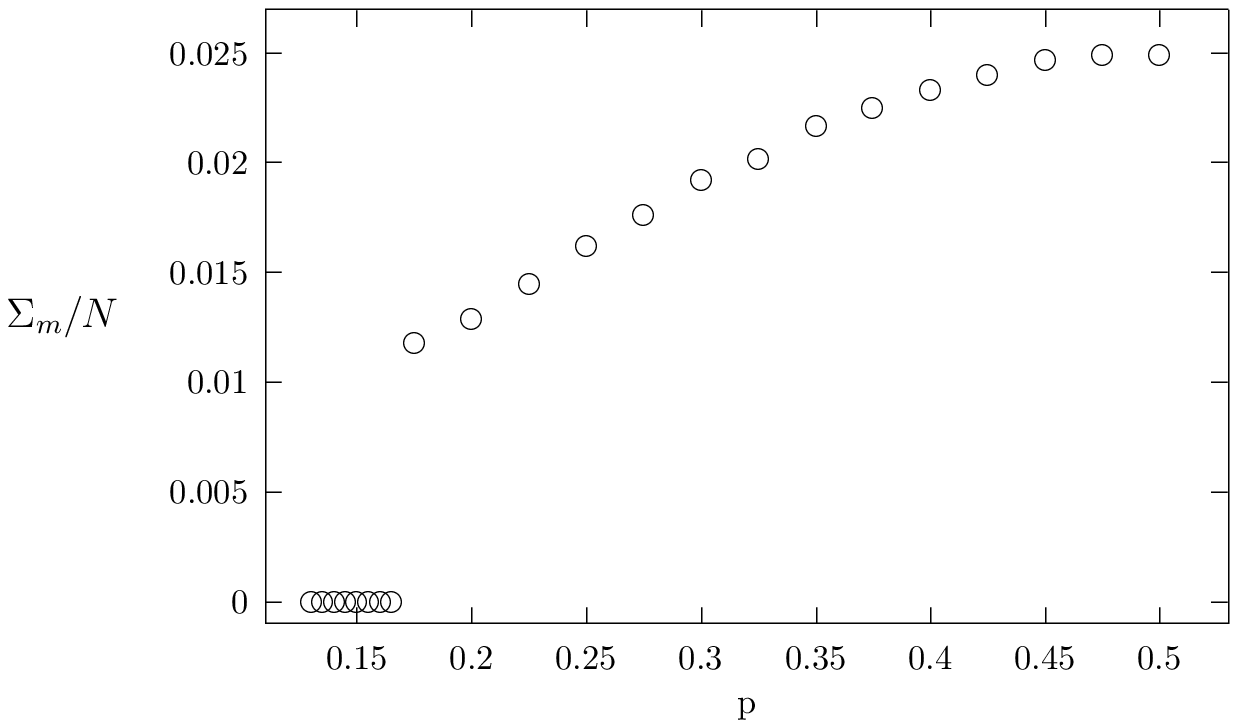}
\caption{Maximum complexity of 3-SAT in terms of $p$. The
parameters are the same as Fig.\ref{f7}. }\label{f8}
\end{figure}

\section{Conclusion}\label{7}
In summery we studied biased random K-SAT problems in which a
variable is negated with probability  $p$. This definition enables
us to go continuously from easy random K-SAT problems to the hard
ones. Certainly this can help us in a better understanding of the
typical complexity of random K-SAT problems. In this paper we gave
the exact solution of 1-SAT case and the full picture of general
K-SAT problems in the replica symmetry approximation. From these
results, which are exact for $K=2$, one can obtain an upper bound
for the critical value of $\alpha_c(p,K)$. We found that
$\alpha_c(p,K)$ has a power law behavior $p^{-\tau_K}$ for
$p\rightarrow 0$ where $\tau_K=K-1$. We studied 3-SAT problems
with the help of numerical solution of the survey propagation
equations and found no replica symmetry breaking transition for
$p<p^* \sim 0.17$.  However in contrast to the tricritical point
of 2+p-SAT problem we found that in both sides of $p^*$ the
SAT-UNSAT transition is discontinuous. This phenomenon dose not
support the current belief that hardness of a problem may stem
from the discontinuous nature of its transition. Certainly it
still demands more studies to have a clear picture of the origins
of typical complexity in these problems.

\acknowledgments We thank M. Mezard for helpful suggestions in
writing this paper. A. R. is grateful to V. Karimipour for the
introduction of this beautiful field to him.

\end{document}